# COMPARATIVE ANALYSIS OF PHENOMENOLOGICAL APPROXIMATIONS OF THE LIGHT CURVES OF ECLIPSING BINARY STARS WITH ADDITIONAL PARAMETERS

I.L.Andronov[1], M.G.Tkachenko[1], L.L.Chinarova[2]

[1] Department «Mathematics, Physics and Astronomy», Odessa National Maritime University, 34 Mechnikov Str., Odessa, 65029, Ukraine; e-mail: tt_ari@ukr.net, masha.vodn@yandex.ua,
[2]Astronomical Observatory of I.I.Mechnikov Odessa National University, 1v Marazliyivska Str., Odessa, 65014, Ukraine; e-mail: lidia_chinarova@mail.ua

*A comparative analysis of the special shapes (patterns, profiles) of the eclipses applied for the phenomenological modeling of the light curves of eclipsing binary stars is conducted. Families of functions are considered, generalizing local approximations (Andronov, 2010, 2012) and the functions theoretically unlimited in a width, based on a Gaussian (Mikulášek, 2015). For an analysis, the light curve of the star V0882 Car = 2MASS J11080308 - 6145589 of the classic Algol - subtype (β Persei) is used. By analyzing dozens of modified functions with additional parameters, it was chosen the 14 best ones according to the criterion of the least sum of squares of deviations. The best are the functions with an additional parameter, describing profiles, which are limited in phase.*

## INTRODUCTION

Phenomenological modeling of light curves of eclipsing binary stars allows get parameters, which are needed for registration of the object in the "General Catalog of Variable Stars" (GCVS) [24], "Variable Stars Index" VSX (http://aavso.org/vsx) and same catalogs. It is the preliminary method for a small fraction of stars, which are then studied using different methods like spectroscopy, polarimetry or multicolor photometry. For most stars, there are no more studies, and the phenomenological modeling remains the main source of information about the object. Classic methods of the variable stars research are described, for example, in [27]. Methods of approximation of symmetric and asymmetric extrema are reviewed in [3].

In most cases, the parameters were determined from the individual parts of the light curve. These include the brightness values at the primary maximum and minimum, and for the eclipsing systems – also at the secondary minimum and maximum, that is indicated in the notes in the GCVS [24]. Becides, for eclipsing systems, there is a necessary parameter - the minimum width D, and the desirable - the duration of the total eclipse phase.

When analyzing total light curves are used as graphic smoothing methods, and approximation of trigonometric polynomial (truncated Fourier series). For the stars EA type (Algol type), number of the parameters becomes very large, which leads to the appearance of visible waves in the light curve (Gibbs effect [26]), and also increase statistical error of the smoothing light curve and the corresponding values at the maximum and minimum.

Moreover, smoothing approximations (including trigonometric polynomial [2,8], and “symmetric” trigonometric polynomial [23]) do not allow to define one of the necessary parameters - width eclipse. Therefore, there is need to administration functions ("special shape"), which would be described statistically optimal eclipse and using a smaller number of parameters. Andronov [4,5] offered approximation “NAV” (“New Algol Variable”), which is applied by us for many stars, not only the type of Algol (EA), but also for two another types – EB (β Lyrae) и EW (W Ursae Majoris)[7, 9, 11, 27]. Approximations based on the Gaussian and it’s improvements, were discussed by Prof. Mikulášek [19,20].

An alternative approach is to physical modeling of light curves, based on the method Wilson–Devinney [31], for the implementation of which various authors have developed relevant programs [12, 13, 16, 17, 22]. However, the necessary parameters for physical modeling (temperature of at least one of the components and the mass ratio), that can be reliably obtained from spectroscopic observations, and they conducted for ~1% known eclipsing binary systems.

Another possibility is to use a simplified physical model, in which the stars assume spherically symmetric, and neglect the effect of limb darkening [25]. This model was used to study and classification of eclipsing binary stars [18], including, for review OGLE [14], and also implemented by us [10].

In this paper, we research the phenomenological modeling modifications to improve the quality of approximation by introducing one or more additional parameters. The work was carried out in the framework of the international project “Inter-longitude astronomy” [1,6] and the national project “Ukrainian Virtual Observatory” [29,30].

## DATA

To illustrate the use of the proposed functions, we used the observation of one of the star Algol type (2MASS J11080308-6145589[21], which recently was named V0882 Car [24]). Of the total phase curve, published in OEJV [21], were used the data in the phase interval [–0.08, 0.08], in which were obtained 120 brightness observations. It allows to study a separate minimum to compare the approximations of the eclipse, while, for the complete study, we use complete phase curves, taking into account the effects of reflection, ellipticity, O'Connell and differences in the profiles of primary and secondary minima of the same width [7, 9, 11, 27].

## THE BASIC FORMULAS

The independent variable is used as the phase. However, the value of the initial epoch, determined some other methods can be somewhat shifted, therefore, for modeling eclipses is necessary to use a phase difference $u = \phi - \phi_0$, where $\phi_0 -$ phase, corresponding to the middle of the contribution, describing the eclipse. Approximation shape of the light curve can be written in a general form as

$$x(\phi) = C_1 - C_2 G(\phi - C_3; C_4; \dots C_m), \qquad (1)$$

Where $C_1 -$ smoothed light value on phase $\phi = C_3$ taking into account the eclipse, $C_2 -$ the amplitude of the eclipse, and function $G$ depends on both the phase and additional parameters describing the pattern (shape) of the eclipse. Of these the most important is $C_4$, which describes the characteristic width minimum. If use the functions $G$, changing only within eclipse, it is convenient to introduce $C_4$, as a half-width of the eclipse, and a dimensionless parameter $\varepsilon = z = \frac{\phi - C_3}{C_4}$. Below we will use variables $u$ and $\varepsilon$ for unlimited profiles for the phase, and $z -$ to the minima with a finite half-width $C_4$, i.e. $-1 \le z \le 1$.

Contrary to the articles, where we defined $C_1$, as an integral average brightness value of the approximation using the trigonometric polynomial (truncated Fourier series) of the second degree excluding contributions of eclipses, in this paper, we redefine $C_1$, as the smoothed value of brightness at the minimum. This function $G(z) = 1 - H(z)$, where $H(z) -$ the function used previously [9,11].

The basic properties of functions $G(z)$: $G(0) = 0$, and, if $|z| > 1, G(z) = 1$. For reasons of symmetry, it is advisable to define the function $G(z) = G(-z) = G(|z|)$, as a symmetric one.

The classical function, which is used, for example, for a first approximation of profiles of spectral lines, widened due to the Doppler effect, is the Gaussian

$$G(u) = (1 - \exp(-|C_4| u^2)). \qquad (2)$$

In this case, $|C_4| = \frac{1}{2\sigma^2}$, where $\sigma -$ characteristic width, which has a sense of the standard deviation in the theory of probability. This function is characterized by only four parameters, what does not allow taking into account the diversity of observed minima profiles.

In Table 1, our best selected approximation are shown, and the Gaussian takes last 14th place according to the criterion of minimal SSE (the sum of squared residuals). For the researched series and function, SSE=0.00736. And although a bad convergence to Gaussian observation is observed in almost all of eclipsing stars, continue to use it [15], apparently due to the popularity of statistics and applicability of approximation in a number of software packages.

To account for the fact that real minimums have a finite duration, Andronov [4,5] proposed to use the function

$$G(z) = 1 - (1 - |z|^{\beta})^{1.5}. \quad (3)$$

Here the power 1.5 corresponds to the theoretical asymptotic behavior of the light curve near the borders of the eclipse. The method was named NAV ("New Algol Variable").

For our example, this method gives SSE=0.00477 (6th place in the ranking). This is only 4% less than the value for the best approximation with a large number of parameters.

Decomposition of this function into Mac-Laurin series

$$G(z) = 1 - (1 - |z|^{\beta})^{1.5} = \frac{3}{2}|z|^{\beta} - \frac{3}{8}|z|^{2\beta} - \frac{1}{16}|z|^{3\beta} - \frac{3}{128}|z|^{4\beta} + \cdots (4)$$

shows, that asymptotically at small $|z|, G(z) \sim |z|^{\beta}$, and this parameter defines the profile of the function. Since we are talking about the vicinities of minimum (at $|z| \approx 0$), than, with classic studies of functions, it is expected that there will be a positive second derivative of the function. In this case, the second derivative is equal to

$$G''(z) = \frac{3}{2}\beta(\beta - 1)|z|^{\beta-2}, \quad (5)$$

which is definitely positive and finite, if $\beta = 2$. If $\beta > 2, G''(z) = 0$, and the minimum is more flat, than the asymptotically parabolic minimum expected for the majority of analytic functions. Parameter $\beta = 1$ corresponds to the "triangular" profile of minimum, since at $z \to 0$, asymptotically $G(z) = \frac{3}{2}|z|$. In this case, the first derivative is discontinuous at $z = 0$. Physically, the minimum limit is $\beta = 1.5$, which corresponds to an instantaneous transition from the entrance to the exit of the eclipse ("instantaneous" total eclipse of a smaller star).

The parameter values $\beta \gg 2$ describe the "flat" minima, or "total eclipses". However, for a significant number of stars, which we studied, the statistically optimal value is $\beta < 2$, which leads to discontinuity of the second derivative and visually a "sharp" profile. However, for these stars, $\beta \geq 1.5$, so the approximation is physically real, although being different from the usual study of analytic functions with $\beta = 2$.

Mikulášek et al. [19] modified the classical Gaussian, also to get the asymptotically power function near the minimum. In our notation:

$$G(u) = (1 - \exp(-|C_4|u^2))^r = (1 - \exp(-\theta))^r, \quad (6)$$

SSE=0.00599 (13th place). So the introduction of an additional parameter $C_5 = r$ leads to an improvement of quality approximation, however, usually concedes to our method NAV.

Decomposition into Maclaurin series for this function gives

$$G(u) = \theta^r \left(1 - \frac{r\theta}{2} + \frac{(3r^2+r)\theta^2}{24} - \frac{(r^3+r^2)\theta^3}{48} + \cdots\right), \quad (7)$$

so asymptotically $G(u) \sim \theta^r = |C_4|^r u^{2r}$, the function similar to that obtained for the method NAV for $\beta = 2r$, and the main difference is near borders of eclipses.

To improve the approximation, without attracting additional parameters, Mikulášek [20] replaced the parabola in the exponent by the hyperbolic cosine

$$G(\varepsilon) = (1 - \exp(1 - \cosh\varepsilon))^r, \quad (8)$$

that corresponds to a 10 place in the ranking of approximations for $r = 1$.

Adding to the existing 5 parameters two more, Mikulášek [20] proposed a modification

$$G(\varepsilon) = 1 - (1 - (1 - \exp(1 - \cosh(\varepsilon)))^{C_5})(1 + C_6\varepsilon^2 + C_7\varepsilon^4), \quad (9)$$

which corresponds to a rating of 8. Thus, the approximation NAV for the series in question is preferred, than those considered [20].

In [11, 9], we considered the modification of functions, which were proposed previously by us [4, 5] and Mikulášek [20]. In this work, a list of modifications was extended, and the best ones are shown in Table 1.
Among them is on the Mikulášek's function:

$$(1 - \exp(|C_6|(1 - \cosh(\varepsilon))))^{C_5}, \quad (10)$$

to which we have added an additional parameter $|C_6|$ (it is absent Mikulášek [20], so it can be considered as equal to unity). Note that this function is the "standard" one for determination of the moments of minima of eclipsing stars at the website *var2.astro.cz* .

More successful was a modification by moving the power index to the parameter of the exponent

$$1 - \exp(-|C_6|(\cosh(\varepsilon) - 1)^{C_5}), \quad (11)$$

The approximation rating is 7 (SSE=0.480).

Obviously, the choice of a hyperbolic cosine instead of the normal parabola is associated with a sharp decrease argument of an exponent at recession from zero:

$$1-\cosh(\varepsilon) = -\frac{\varepsilon^2}{2} - \frac{\varepsilon^4}{24} - \frac{\varepsilon^6}{720} - \frac{\varepsilon^8}{40320} - \cdots \quad (12)$$

We also tried an approximation, limited in phase, taking into account decomposition in even powers of the argument

$$G(z) = \frac{(1-\exp(-|C_5|z^2-|C_6|z^4-|C_7|z^6))^{C_8}}{(1-\exp(-|C_5|-|C_6|-|C_7|))^{C_8}}, \quad (13)$$

however, it was less successful (rating 11, SSE=0.00544). Signs of absolute values show that the parameters are positive.

Thus, an unlimited approximations using exponential and hyperbolic cosine, has ratings 7 and larger, so worse than of the original algorithm NAV.

However, we tried to improve the method NAV, and modify the eclipse profile, limited in phase. Six functions showed the same results on SSE within a statistically insignificant difference in 1.5%. In order to improve the rating, it is an approximation with the "total eclipse" shape

$$G(u) = \begin{cases} 0, & \text{if } |u| \le C_6 \\ 1-(1-((|u|-|C_6|)/(C_4-|C_6|))^{C_5})^{1.5}. & \text{if } C_6 < |u| < C_4 \end{cases}, \quad (14)$$

the approximation with an additional power index $C_6$ (previously treated equally 1.5)

$$G(u) = 1-(1-|z|^{C_5})^{C_6}, \qquad |z| \le 1. \quad (15)$$

series in even powers (which can be obtained by decomposition in the series of exponent with modified index by us), raised to the power

$$G(z) = (|1-C_6-C_7-C_8|z^2+C_6z^4+C_7z^6+C_8z^8)^{C_5}, \qquad |z| \le 1. \quad (16)$$

An excellent result gives the Mikulášek's function, which we modified to be limited in phase

$$G(z) = \frac{(1-\exp(1-\cosh(z)))^{C_5}}{(1-\exp(1-\cosh(1)))^{C_5}}, \quad |z| \le 1, \quad (17)$$

However, the best result for the studied series shows the method NAV with modified argument with additional parameter $C_6$ in range (-1,1):

$$G(z) = 1-(1-(|z|+C_6|z|(1-|z|))^{C_5})^{1.5}, \quad |z| \le 1. \quad (18)$$

## Theoretical class of functions

Figure 2 shows class of approximations for the basic methods, modifications of which have been studied in this paper. At the top, is shown the function limited in phase (3), proposed by Andronov [4, 5], and at the bottom – the "unlimited" function (10) by Mikulášek [20]. For comparison of the profile near the center of the eclipse, we accepted values $\beta = 2r$ for the same data set.

The main difference between these classes – in limited or unlimited in phase theoretical profiles, and near the center of the eclipse, profiles are asymptotically of the same power.
Figure 3 shows the class of approximations for the best possible modification (18). The addition of an additional parameter has reduced the value of SSE by 2.5%, as compared to the original formula (3). Thus, the function seems to be most perspective for improving approximation of the eclipse profile. The natural physical limitation for this parameter is the interval (-1,1). Otherwise, the function becomes no more monotonic while changing from the center to the edges of the eclipse.

## THE DISCUSSION OF THE RESULTS

In the article [4,5] we discussed various functions for the approximation. Later Mikulášek [20] introduced several new functions, however, they were still formally infinitely wide. Those phenomenological approximations with a larger number of parameters describe a curve with a smaller standard deviation. However, a common problem due to increase of the number of parameters, is that the functions do not make an orthogonal basis, with a corresponding strong deterioration in the accuracy of determination of even the previous parameters. We conducted a comparison of some modifications of the NAV method and the methods based on the Gaussian [9,11]. In this work, a list of approximating functions is significantly extended, however, it is about one third of the modifications carried out by us. To reduce the influence on the approximation of the part of the light curve outside the eclipse, we have extracted observations of the star V0882 Car= 2MASS J11080308-6145589 [21] within the range close to the minimum from the phases -0.08 to 0.08, the total number is 120. To approximate using the method of nonlinear least-squares, we have used a shareware program WinCurveFitv1.1.2 (Kevin Raner Software), which allows to determine (at maximum) 8 parameters.

The results are shown in Table 1, corresponding smoothing curves – in Figure 1 in order of degradation of the quality of approximations.

As a primary test-function for ranking the approximations, the sum of squared residuals was used.

The best approximation method was NAV, to which was added a correction parameter. However, very little difference between the test-functions does not allow a conclusion about undeniable advantage of any approximation as compared to several practically equivalent ones.

The method of Mikulášek gives almost the same quality of approximations, but, as mentioned above, formally eclipse width, where the function, describing the profile, intersects zero, is equal to infinity. Note, that quick methods for the period search and automatic classification of stars commonly use approximations, which have significantly worse agreement between the theory and the observations, however, calculated relatively fast. These include the classic pre-study method of Algols, when the light curve was divided into "eclipse" (loss of brightness below a certain level) and a "part outside the eclipse", and the approximations of minima using a triangle or a parabola. We also analyzed these data approximations, but, they are characterized by significantly worse values of the test-function SSE, and that's why are suitable only for preliminary (low accuracy) estimates of the parameters.

## CONCLUSION

Discussed functions and their modifications showed a natural improvement of the quality of approximation with increasing number of parameters using the criterion of the sum of squared residuals SSE, however, the effect of adding a parameter significantly different for various initial functions.

Among 14 best functions, the best approximation, without phase limits, has a rating of only 7. The best six corresponds to limited width of minimum (what is expected in the physical sense), either for modification of functions (3) (Andronov [4,5]), or for the function (10) (Mikulášek [20]). Modifications of both classes asymptotically have power dependence near the center of the eclipse, differing on the edges.

The best for the test series of observations is the approximation (18) with «phase distortion», however, the statistical significance of the parameter must be determined for each series of observations separately.

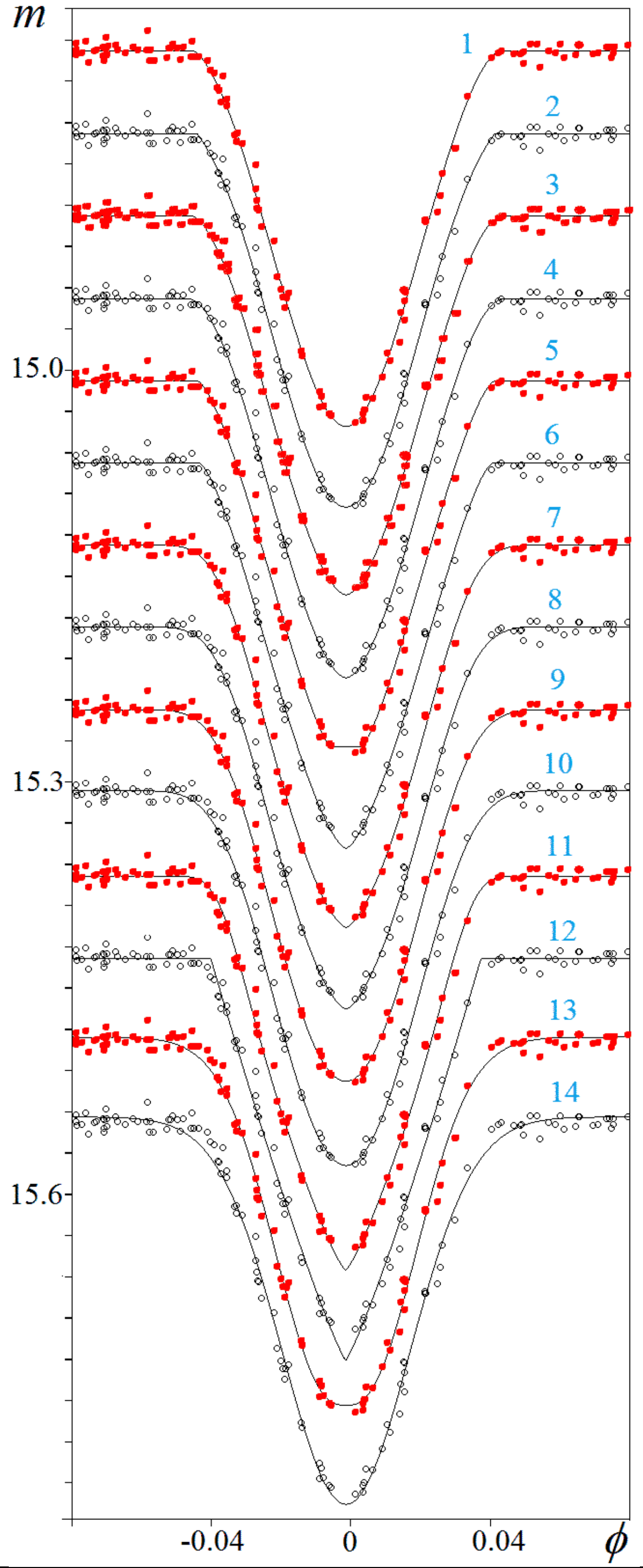

Fig. 1. Approximations of the light curve near the brightness minimum of the star V0882 Car= 2MASS J11080308-6145589 by various functions. The numbering of curves corresponds to that shown in Table 1.

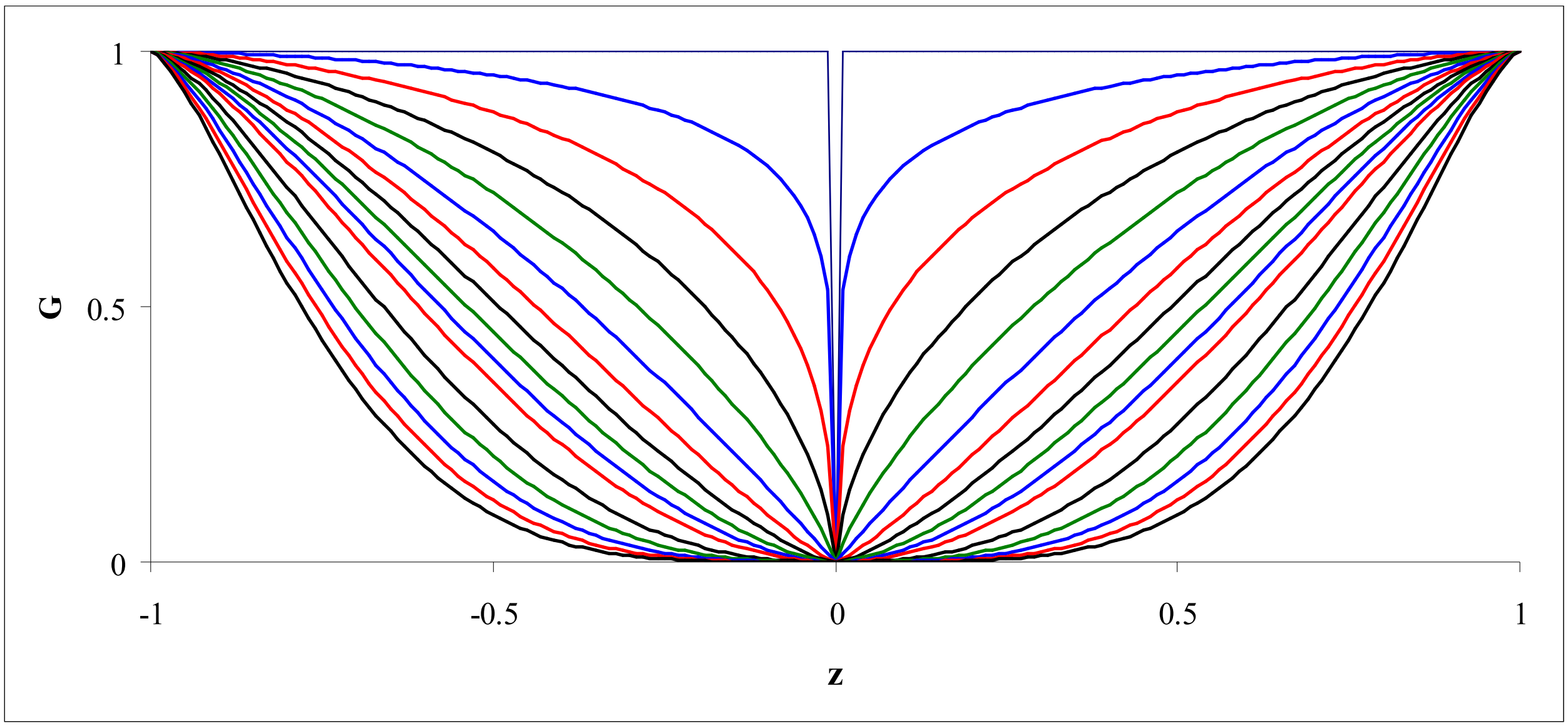

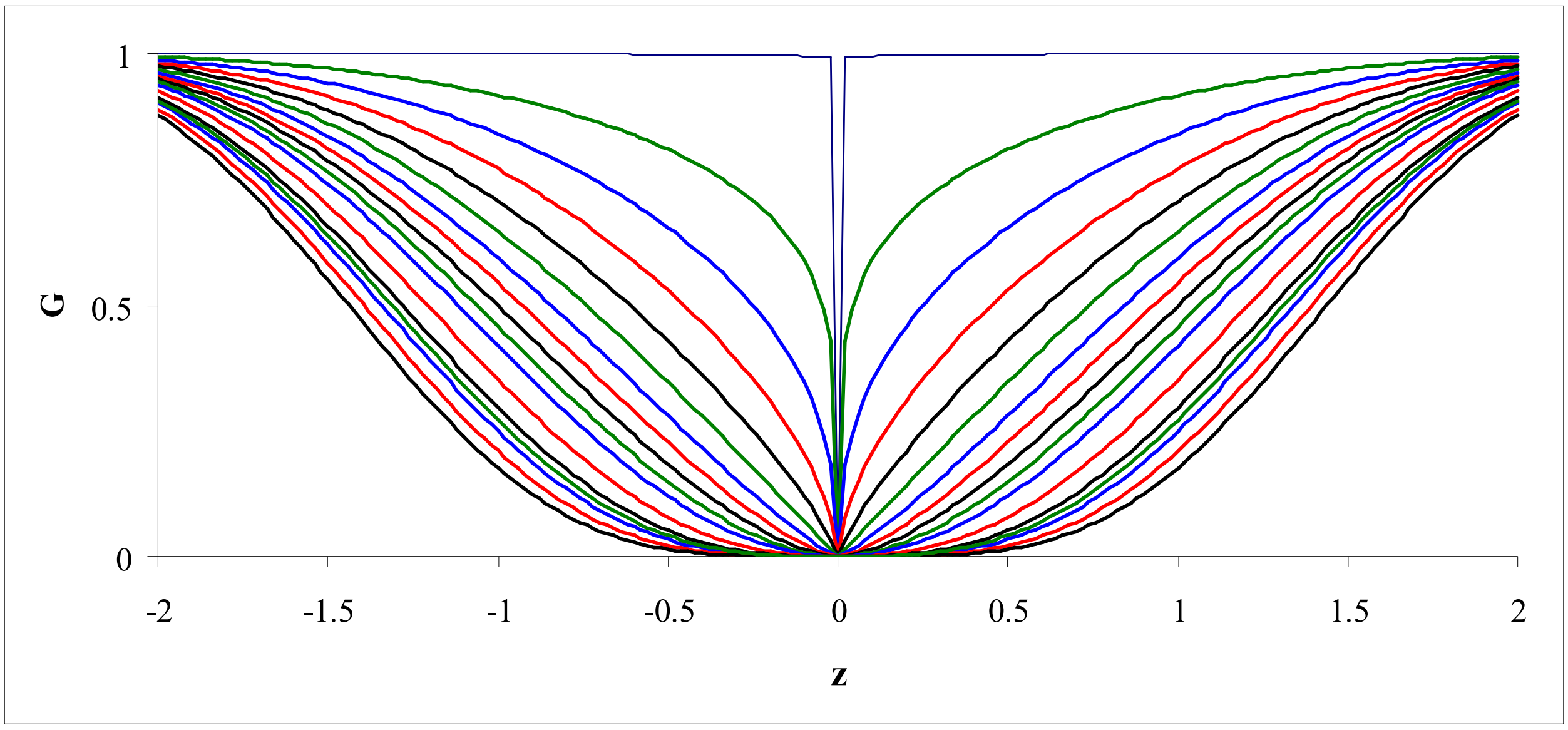

Fig. 2. a) Influence of the parameter ($\beta = C_5$ in equation (3)) for values $\beta = 2r$ and $r$= 0.001, 0.1, 0.2, 0.3, 0.4, 0.5, 0.6, 0.7, 0.8, 0.9, 1, 1.2, 1.4, 1.6, 1.8, 2 (width increases); б) Influence of the parameter $r(= C_5$ in equation (10)) on the form of light curves for the same set of values, as in Figure. (a).

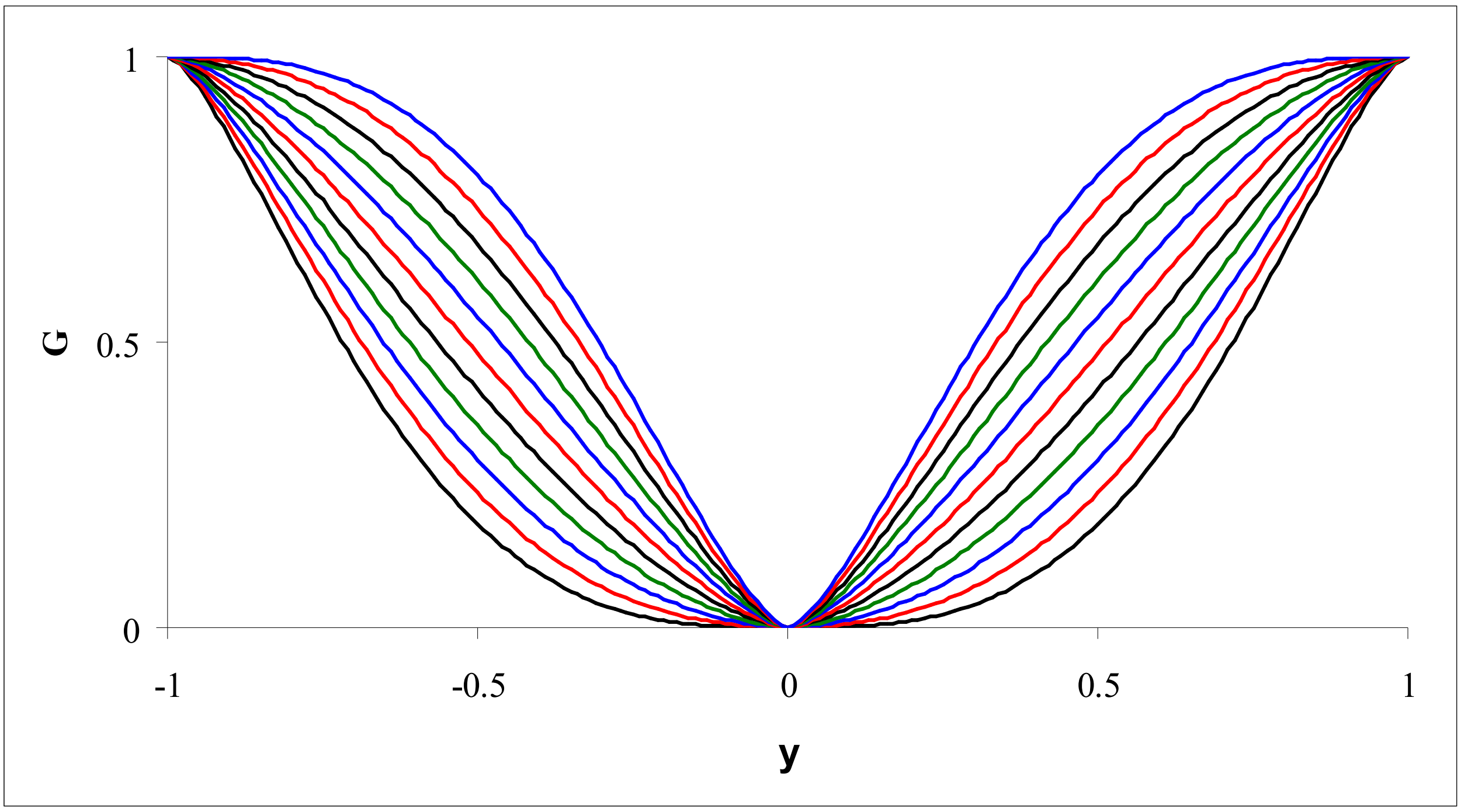

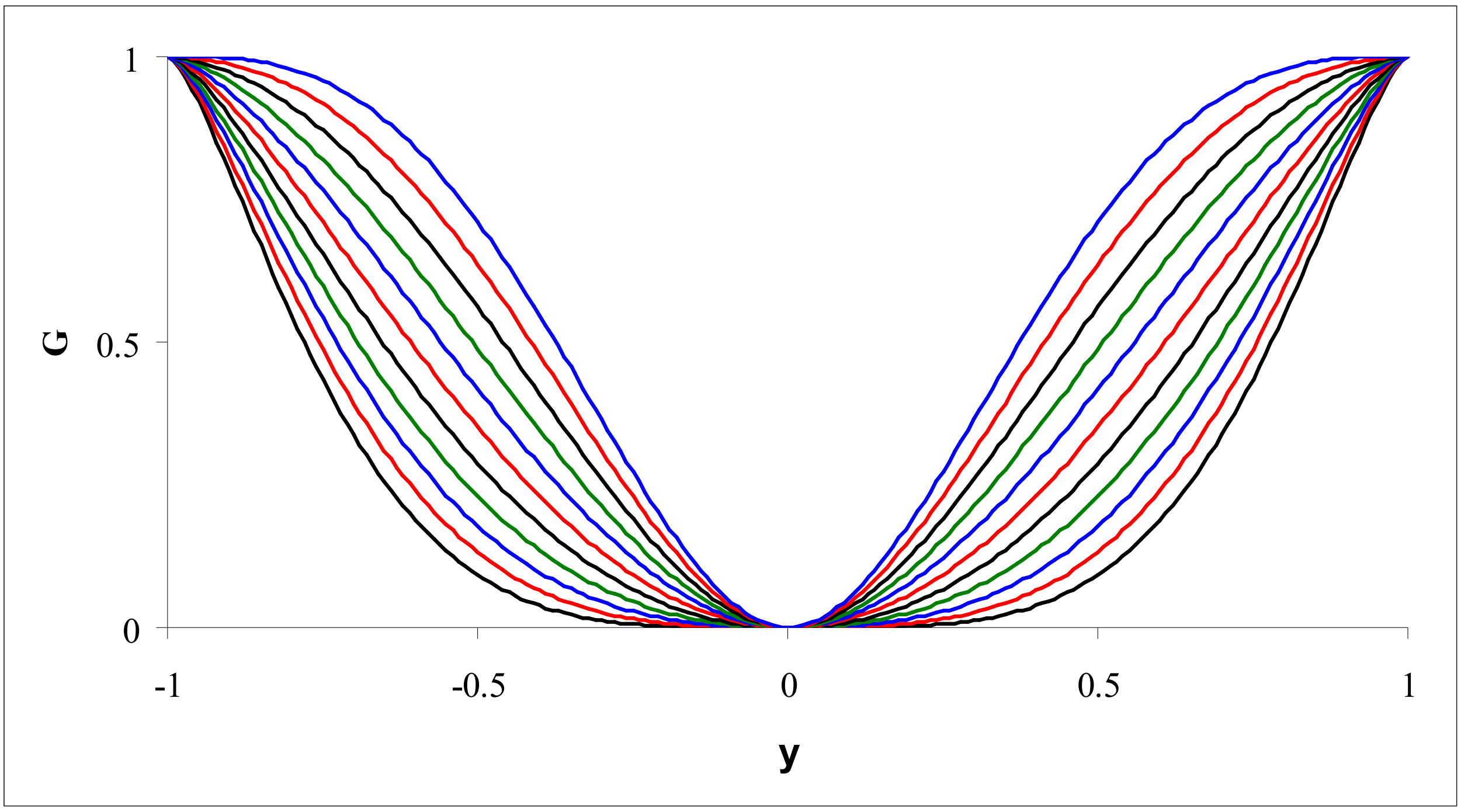

Fig. 3. Influence of the parameter $C_{13}$ on profile of minimum at a fixed value: a) $\alpha$=1,5; b) $\alpha$=2.

Table 1. Characteristics of the best approximations.

| № | Equation | SSE | $\sigma[C_3]$ | $X_C(C_3)$ | $\sigma[X_C]$ | m |
|---|---|---|---|---|---|---|
| 1. | $1-(1-(\lvert z\rvert+C_6\lvert z\rvert(1-\lvert z\rvert))^{C_5})^{1.5},\ \lvert z\rvert\le 1$ | 0.00459 | -0.00139 | 15.0412 ±0.0034 | 0.00142 | 6 |
| 2. | $\frac{(1-\exp(1-\cosh(z)))^{C_5}}{(1-\exp(1-\cosh(1)))^{C_5}},\ \lvert z\rvert\le 1$ | 0.00462 | -0.00139 ±0.00012 | 15.0394 ±0.0080 | 0.00129 | 5 |
| 3. | $(\lvert 1-C_6-C_7-C_8\rvert z^2+C_6z^4+C_7z^6+C_8z^8)^{C_5},\quad \lvert z\rvert\le 1$ | 0.00462 | -0.00139 ±0.00012 | 15.0437 ±0.0035 | 0.00166 | 8 |
| 4. | $1-(1-\lvert z\rvert^{C_5})^{C_6},\qquad \lvert z\rvert\le 1$ | 0.00462 | -0.00139 ±0.00012 | 15.0435 ±0.0032 | 0.00142 | 6 |
| 5. | $\begin{cases} 0, \text{if} \lvert u\rvert \le C_6 \\ 1-(1-((\lvert u\rvert-\lvert C_6\rvert)/(C_4-\lvert C_6\rvert))^{C_5})^{1.5}, \text{if } C_6<\lvert u\rvert < \end{cases}$ | 0.00465 | -0.00139 ±0.00012 | 15.0338 ±0.0040 | 0.00143 | 6 |
| 6. | $1-(1-\lvert z\rvert^{C_5})^{1.5},\qquad \lvert z\rvert\le 1$ | 0.00477 | -0.00139 ±0.00012 | 15.0485 ±0.0031 | 0.00131 | 5 |
| 7. | $1-\exp(-C_6(\cosh(\varepsilon)-1)^{C_5})$ | 0.00480 | -0.00139 ±0.00012 | 15.0458 ±0.0031 | 0.00145 | 6 |
| 8. | $1-(1-(1-\exp(1-\cosh(\varepsilon)))^{C_5})(1+C_6\varepsilon^2+C_7\varepsilon^4)$ | 0.00487 | -0.00139 ±0.00012 | 15.0455 ±0.0087 | 0.00159 | 7 |
| 9. | $1-\exp(C_5(1-\cosh(\varepsilon)))$ | 0.00500 | -0.00142 ±0.00012 | 15.0377 ±0.0035 | 0.00135 | 5 |
| 10. | $1-\exp(1-\cosh(\varepsilon))$ | 0.00509 | -0.00140 ±0.00013 | 15.0393 ±0.0034 | 0.00121 | 4 |
| 11. | $\frac{(1-\exp(-\lvert C_5\rvert z^2-\lvert C_6\rvert z^4-\lvert C_7\rvert z^6))^{C_8}}{(1-\exp(-\lvert C_5\rvert-\lvert C_6\rvert-\lvert C_7\rvert))^{C_8}},\ \lvert z\rvert\le 1$ | 0.00544 | -0.00146 ±0.00013 | 15.0550 ±0.0143 | 0.00180 | 8 |
| 12. | $1-(1-z^2)^{C_5},\qquad \lvert z\rvert\le 1$ | 0.00581 | -0.00143 ±0.00014 | 15.0609 ±0.0184 | 0.00145 | 5 |
| 13. | $(1-\exp(-\lvert C_4\rvert u^2))^{C_5}$ | 0.00599 | -0.00143 ±0.00014 | 15.0334 ±0.0053 | 0.00147 | 5 |
| 14. | $(1-\exp(-\lvert C_4\rvert u^2))$ | 0.00736 | -0.00134 ±0.00016 | 15.0456 ±0.0050 | 0.00145 | 4 |